\documentclass[runningheads]{llncs}
\usepackage[T1]{fontenc}
\usepackage[hidelinks]{hyperref}
\usepackage{graphicx}
\usepackage{xcolor}
\usepackage{cite}
\usepackage{listings}

\usepackage{paralist}
\newcommand{\toolname}{{\sc ARCeR}}
\begin{document}
\title{\toolname{}: an Agentic RAG  for the Automated Definition of Cyber Ranges}
\titlerunning{\toolname{}: an Agentic RAG for Cyber Ranges}

\author{
    Matteo Lupinacci\inst{1}\orcidID{0009-0000-2356-398X}\and
    \\
    Francesco Blefari\inst{1,2}\orcidID{0009-0000-2625-631X}\and
    \\
    Francesco Romeo\inst{1,2}\orcidID{0009-0006-3402-3675}\and
    \\
    Francesco Aurelio Pironti\inst{1}\orcidID{0009-0003-3183-2977}\and
    \\
    Angelo Furfaro\inst{1}\orcidID{0000-0003-2537-8918}
}
\authorrunning{Lupinacci et al.}

\institute{
University of Calabria, 87036, Italy \\
\email{\{matteo.lupinacci,francesco.blefari, francesco.romeo,  francesco.pironti, angelo.furfaro\}@unical.it}\\
\and
IMT School for Advenced Studies, Lucca, 55100, Italy
}

\maketitle
\begin{abstract}
The growing and evolving landscape of cybersecurity threats necessitates the development of supporting tools and platforms that allow for the creation of realistic IT environments operating within virtual, controlled settings as Cyber Ranges (CRs).  CRs can be exploited for analyzing vulnerabilities and  experimenting with the effectiveness of devised countermeasures, as well as  serving as training environments for building cyber security skills and abilities for IT operators.
This paper proposes \toolname{} as an innovative solution for the automatic generation and deployment of CRs, starting from user-provided descriptions in a natural language. \toolname{} relies on the Agentic RAG paradigm, which allows it to fully exploit state-of-art AI technologies. Experimental results show that \toolname{} is able to successfully process prompts even in cases that LLMs or basic RAG systems are not able to cope with.   Furthermore, \toolname{} is able to target any CR framework provided that specific knowledge  is made available to it. 
\end{abstract}

\section{Introduction}
\label{sec:intro}
In recent years, the increasing number of threats targeting IT systems has led to a significant focus in the cybersecurity research community on the development of Cyber Ranges (CR) as fundamental tools to train IT professionals in facing cyber threats and attacks. Typically, setting up a CR involves defining its infrastructure and software in a configuration file. A CR platform admin (such as cybersecurity instructors) has to manually design, build and deploy custom scenarios  by writing these files, an activity which is both time consuming and prone-to-error.  
In addition, the complexity of building and maintaining real-world scenarios, both  in enterprise and educational settings, is challenging. 

In the last few years, CR frameworks, as indicated in~\cite{Grimaldi_2024} and~\cite{Crate_paper}, have become a widespread solution in assisting the CR development life-cycle. 
Several frameworks have been proposed, each with distinct features and objectives, that mainly focus on one (or both) of the following two main aspects:
\begin{inparaenum} [(i)]
    \item enhance the generation of increasingly sophisticated training scenarios that accurately reflect real-world threats and
    \item  reduce the time and resources necessary for the configuration and execution of CR instances (see Section~\ref{sec:related}). The latter means that the framework automatically provides the necessary infrastructure and establishes connections between machines according to their network topology without the intervention of the instructor. 
\end{inparaenum}

The use of machine learning (ML) to improve CR deployment is explored in~\cite{OpenStack_based_CR}, where a system based on the OpenStack cloud platform is presented to automate CR deployment. This system uses ML to classify VMs to reduce the cost of manual selection. With that platform, the instructor only needs to provide a YAML description of a virtual environment that is parsed to complete the automated deployment.

Among the more innovative approaches, the work by~\cite{LLM_for_CR} stands out as the first to propose the application of LLM to the CR domain. Their approach transforms the well known and inherent in LLMs ‘hallucination’ problem into a potential advantage, allowing the creation of complex scenarios that push the boundaries of traditional cybersecurity training.

To the best of our knowledge, this is the first work which proposes the use of Large Language Models (LLMs) and Agentic Retrieval-Augmented Generation (Agentic RAG)~\cite{Singh2025} systems for the automated definition and deployment of training scenarios compatible with multiple CR platforms.

The devised approach aims to simplify this process by generating CR from natural language descriptions. The ability of Agentic RAGs to plan, use tools external to the LLM, dynamically adapt to responses from the environment, and perform multiple retrieval steps from external-supplied documents enables them to reduce the costs required for fine-tuning an LLM and adapt to multiple CR instantiation platforms.

Furthermore, the use of augmented knowledge empowers the agent to generate valid configuration files for a given CR platform. This represents a significant advance compared to previous works~\cite{OpenStack_based_CR,LLM_for_CR,CyExec*_CR} that employed specific techniques for the automatic generation of randomized scenarios. In summary, the main contributions of this paper are:
\begin{itemize}
    \item the proposal of a novel approach based on Agentic RAG systems for the automated instantiation of CR from natural language text descriptions of the infrastructure that is compatible with multiple CR platforms;
    \item the support for the autonomous generation of self-devised CR scenarios leveraging Agentic RAG systems;
    \item the design and implementation of \toolname{}, the first Agentic RAG system for CR definition and deployment (this tool will be made available open-source to foster future research in this area).
     
    \end{itemize}
The effectiveness of \toolname{} has been assessed by testing it with CyRIS~\cite{CyRIS} a well-known CR framework.
It has been observed that while the generation of files based on specific knowledge can generally be accomplished with a pure RAG system, employing an Agentic RAG can enhance the accuracy and integrity of the output while maintaining equivalent complexity.

The remainder of this paper is organized as follows. Section~\ref{sec:back} provides the basic background on Cyber Ranges and (AI) Agents. Section~\ref{sec:ragagents} presents the proposed approach to the automatic instantiation, generation and deployment of CR using Agentic RAG. Section~\ref{sec:casestudy}  presents a case study based on CyRIS. Section~\ref{sec:results} discusses the achieved results. Section~\ref{sec:related} overviews the related works and Section~\ref{sec:concl} draws the conclusions.

\section{Background}
\label{sec:back}

\subsection{Cyber Ranges}
\label{subsec:cr}
The US National Institute of Standards and Technology (NIST) defines Cyber ranges as {\it interactive, simulated representations of an organization’s local network, system, tools, and applications that are connected to a simulated Internet level environment. They provide a safe, legal environment
to gain hands-on cyber skills and a secure environment for product development
and security posture testing}~\cite{NIST_CR_1}. 
Cyber ranges are controlled and interactive virtual environments that are utilized for efficient and secure cybersecurity training, as well as for the secure and controlled emulation of new real-world attacks and malware. For this reasons, they can be considered as contemporary battlefields for cybersecurity. Moreover they offer trainees a context for augmenting their cybersecurity skill set through hands-on activities (such as the real-time analysis frameworks for CRs as proposed in~\cite{blefari2024towardlogbased, romeo2025unvealing}).

A CR comprises an IT infrastructure and a suite of selected security features. The infrastructure can include, but is not limited to, machines, networks, storage, and software tools. Security functionalities include the capability to replicate cyberattacks and execute malware in a controlled environment. The orchestration layer of the CR coordinates the diverse technology and service components of the underlying infrastructure and provides isolation from other resources on the host systems. This isolation enables the simulation of complex scenarios without compromising live production systems. A CR may also integrate a Learning Management System (LMS) that allows both instructors and trainees to track and measure progress through a defined training curriculum~\cite{cybsecguide:2023}.

The utilization of CRs has undergone a substantial transition, with a shift from their initial adoption by military and government agencies to their current application by a wide array of businesses and organizations including bug-bounty hunters, researchers and students. Notably, CRs have found application in allowing in a secure way, the dynamic analysis of malware utilized in targeted attacks, where the execution of the malicious code is necessary to determine its purpose. 

According to the NIST, there are four main categories of CRs: (i) simulations, (ii) overlay, (iii) emulation, and (iv) hybrid ranges. These distinctions assume particular significance when aligned with the specific use case of an individual or organization \cite{NIST_CR_2}.

\textit{Simulations} ranges entail the establishment of a synthetic network environment that simulates the behavior of real network components within virtual instances (VMs). VMs are used to mimic specific servers or network of various infrastructures. They offer the advantage of quick reconfiguration because they use standardized templates. However, the fidelity of the exercise increases as the simulation closely matches the target infrastructure. Nevertheless, the unpredictable and unrealistic latency and jitter in network performance are factors that can impact the overall realism of the simulation so they should be limited to the least possible extent. 

\textit{Overlay} ranges provide a higher level of fidelity as they directly utilize the actual network infrastructure. However, this increased fidelity comes with notable costs for hardware and the potential risk of compromising the underlying network infrastructure. This type of CRs are often established as global testbeds for research and experimentation.

\textit{Emulation} is an approach to CR generation that transforms the physical infrastructure into the cyber range itself. It provides closed-network environments that consist of multiple interconnected components and includes traffic generation that emulates various protocols, source patterns, traffic flows and attacks. Emulation gives trainee the authentic experience, rather than pre-programmed actions. A notable example of the use of emulation in CR is the National Cyber Range (NCR) \cite{NCR_USA}. 

\textit{Hybrid} ranges are formed through a customized combination of any of the previously mentioned types that suits specific requirements. A prominent example of Hybrid ranges is the \textit{European Future Internet Research \& Experimentation (FIRE)} project. \cite{FIRE_paper}\\

\subsection{Agents}
An Agent \cite{wooldridge:2009} is defined as a computer system situated in an environment that is capable of acting autonomously in its context in order to reach its delegated objectives. 

\begin{figure}[htb]
    \centering
    \includegraphics[width=0.75\linewidth]{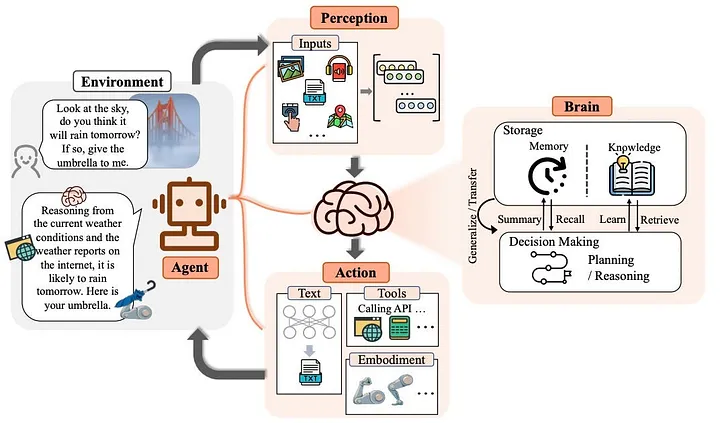}
    \caption{AI  agent structure~\cite{LLM_Agents_survey}}
    \label{fig:AI_agent_schema}
\end{figure}

\textit{Autonomy} means the ability and requirements to decide how to act to achieve a goal. An agent that can perceive its environment, react to changes that occur in it, take the initiative, and interact with other systems (like other agents or humans) is called an intelligent agent or AI Agent (Fig.~\ref{fig:AI_agent_schema}). Another core concept of AI agents is the \textit{memory}. Effective memory management improves an agent's ability to maintain context, learn from past experiences, and make more informed decisions over time. 

As pointed out in~\cite{LLM_Agents_survey}, the emergence of LLMs represents another moment of progress in the realization of AI agents. The substantial advancements made in this direction have culminated in the emergence of LLM agents. In particular, LLM agents use LLMs as reasoning and planning cores to decide the control flow of an application while maintaining the characteristics of traditional AI agents. LLM agents enhanced the LLMs by allowing them to invoke external tools to solve specific tasks, such as mathematical calculations or code execution. In the end, the LLM can decide whether the generated answer is sufficient or if more work is needed.

\subsection{Agentic RAG}
Despite their flexibility, general purpose LLMs, and the related LLM agents, often lack the domain-specific knowledge required to solve particular and non-trivial tasks. Such a problem could be solved by re-training or fine-tuning the model; however, the cost of these operations is not negligible. 

A possible alternative approach is based on the Retrieval-Augmented Generation (RAG)~\cite{RAG} paradigm where the LLM, in pursuing its goal, cooperates with two external components: (i) a source of domain-specific knowledge (i.e. external-supplied documents) and (ii) a  {\it retriever} which is in charge to search for relevant information in the external knowledge base and to augment the context of the LLM. This operation provides the model what is needed to answer complex questions. In RAG systems, external data are loaded and divided into chunks of the appropriate size. The chunks are then converted into vector representations and stored in data structures for future use. The functioning of a RAG system is typically structured as follows:
\begin{inparaenum}[(i)]
\item the user provides a query to the system;
\item the retriever converts the query into a vector representation and performs a match with the stored embeddings, fetching the most relevant chunks;
\item the original query is enhanced with the fetched chunks and passed to the LLM;
\item the context-aware results produced by the LLM are returned to the user.
\end{inparaenum}

An agent system that takes advantage of the RAG paradigm is often referred to as \textit{Agentic RAG}. Many state-of-the-art frameworks, such as LangChain~\cite{LangChain}, LlamaIndex~\cite{LlamaIndex}, and Langdroid~\cite{Langdroid}, provide easy-to-use interfaces to create custom Agentic RAGs.

\section{\toolname{}: Agentic RAG for Cyber Ranges}
\label{sec:ragagents}
Most CR platforms receive as input for CR generation and deployment a \textit{description file} containing information about the host running the range and the CR characteristics (e.g., the virtual machines that make up the range and the network topology) with the corresponding image files. Description files are written in specific formats  such as  YAML~\cite{YAML_site} and, depending on the CR platform and range specification, they can become very complex to write. Traditional CR instantiation methods require manual configuration. 

\toolname{} leverages Agentic RAG systems to create description files from a high-level textual description of the desired specifications and then for automatic CRs deployment on remote servers. More in detail, the primary strength of \toolname{} lies in its user-friendliness for instructors, who can effortlessly create a training environment by simply expressing the desired characteristics in natural language, without the necessity of concern for the underlying framework's specific syntax requirements. This feature makes CR management accessible to users with different levels of expertise. It allows effective use by enabling advanced attack and defense configurations, adjustable difficulty levels, and full infrastructure customization to meet specific training needs in a fully automated manner. 

\toolname{} dynamically and automatically adapts to different CR framework simply by changing the set of documents provided as external knowledge. This ensures a high level of flexibility in that it removes any dependence to specific platforms. This means that alterations in the field, such as the introduction of new configuration patterns or support for novel scenarios, do not necessitate modifications to \toolname{} logic. Only a revision of the reference documents is required, thereby facilitating maintenance and adaptation to emerging requirements. 

Furthermore, by eliminating the need to fine-tune a model for each framework, \toolname{} approach dramatically lowers computational and development costs, making CR generation more affordable, fast, and adaptable to heterogeneous scenarios. Finally, should more efficient LLMs emerge in the future, \toolname{} can be upgraded without the need for a complete system rebuild. These characteristics make the Agentic RAG perfectly suited to the ever-changing context of CRs. 

This work demonstrates (see Section~\ref{sec:results}) that, while pure RAG systems can indeed address the task of generating responses based on specific documentation with sufficient effectiveness, the Agentic RAG approach of \toolname{} leads to enhanced performance. 

\subsection{\toolname{} schema}
\label{sec:schema}
\begin{figure}[ht]
    \centering
    \includegraphics[width=1\linewidth]{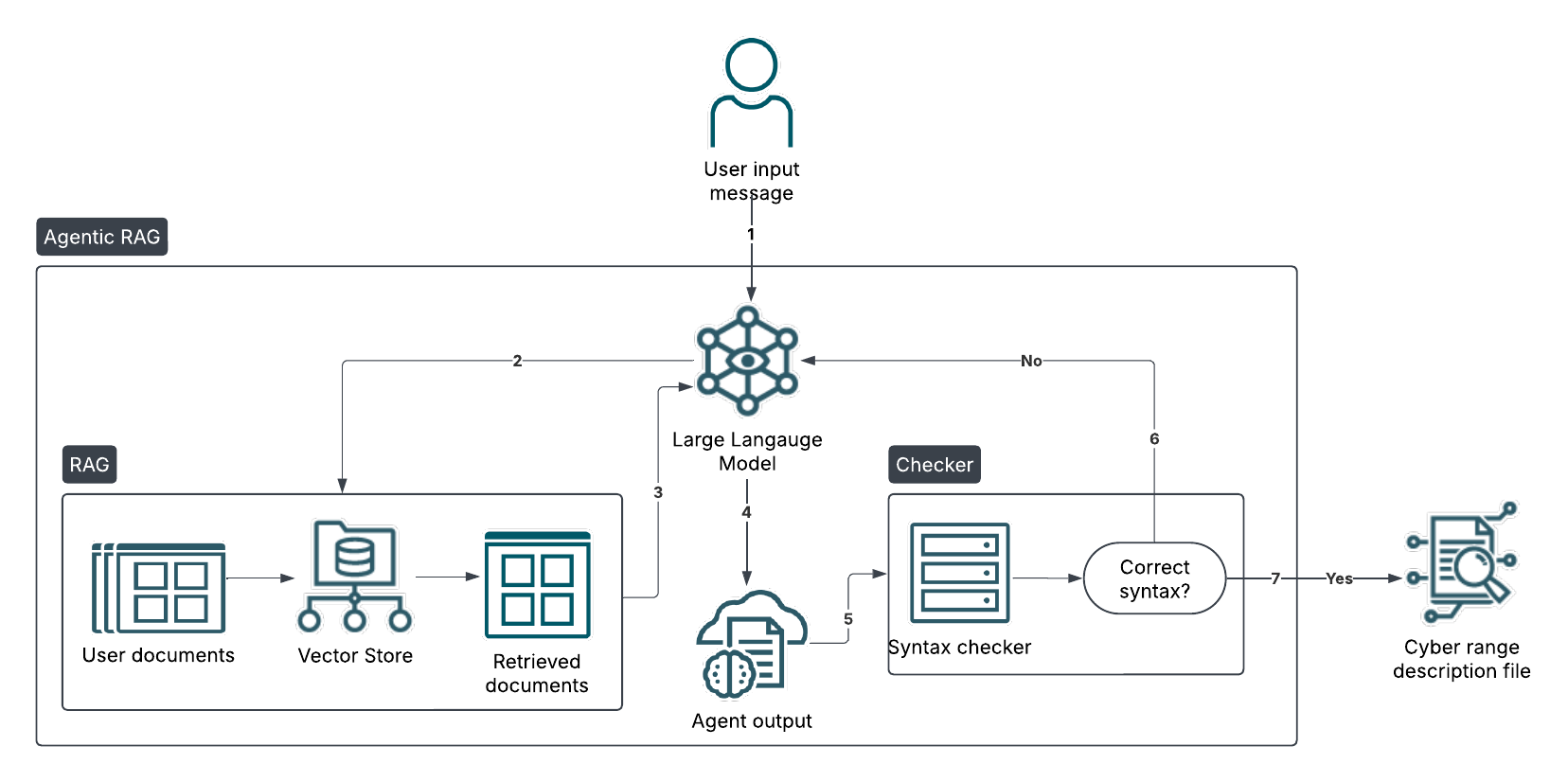}
    \caption{Overall approach schema}
    \label{fig:approach}
\end{figure}

The overall architecture and the operation of \toolname{} are illustrated in Fig.~\ref{fig:approach}. The Agentic RAG structure of \toolname{} includes an LLM that operates as a reasoning engine and two external tools: a \textit{RAG subsystem} and a \textit{Checker Tool}. Finally, being the AI agent a stateless system, it is unable to recall previous interactions with the user. In contrast, \toolname{} has effective memory management, thereby conferring on the instructor the flexibility to modify CR features at a subsequent time. 

The \textit{Checker Tool} allows \toolname{} to check the syntax of the generated output during runtime and to perform self-correction of errors if they occur.

The \textit{RAG subsystem} is in charge of completing the retrieval phase of a RAG system using the Maximal Marginal Relevance~\cite{MMR_metric} (MMR) technique, which aims to mitigate redundancy in the extracted chunks from the vector store. This is achieved by selecting the most relevant documents for the query, ensuring those that differ the most from each other, thereby providing the Agentic RAG with a substantial amount of relevant knowledge. Specifically, 20 chunks are initially extracted, and after filtration, only the 8 most relevant ones are transmitted to the LLM. To achieve an optimal balance between relevance and diversity,  the specific parameter  {\tt lambda\_mult}  was set to 0.5~\cite{MMR_metric}.
\vskip-10pt
\vskip-10pt
Given \toolname{}'s support to interact with disparate CR frameworks it employs different vector stores to separately save the embeddings' chunks documents related to different platforms. All documents (e.g., framework usage guides and example configuration files) are stored in a database containing a folder whose name coincides with the name of the specific CR framework. The RAG subsystem exploits format-specific document loaders (e.g., for PDF and YAML) to import them. 

In more detail, the steps performed by \toolname{} are the following: 
\begin{enumerate}[(1)]
\item The user prompt is fed to the LLM which parses it and evaluates whether it is able to handle it directly or if it needs to call an external tool. In the case it needs to resort to the RAG subsystem it produces a new specific query to be sent to the RAG subsystem. The RAG subsystem identifies the suitable vector store and then it searches for relevant chunks using the MMR technique. On the basis of the additional knowledge contained in these chunks, the LLM produces an initial potential output to the user's query. 
\item The output is sent to an external server that exposes an API to the CR platform component responsible for checking the syntax of the description file. The verification step is a critical component of the overall process as it ensures the integrity of the generated output. 
\item If the above step is successfully completed, the configuration file is created and subsequently returned to the user. In the event that errors are detected, they are automatically sent back to \toolname{} for self-correction and the process starts again from step (1).
\item At the end of the description file generation phase, \toolname{} asks the user whether they want to proceed  with the actual instantiation of the CR. If the user agrees, it directly executes the appropriate commands on the host running the CR framework.
\end{enumerate}

\subsection{Implementation details}
The current implementation of \toolname{}  uses Anthropic's \textit{Claude~3.7~Sonnet}~\cite{Claude_3.7} as LLM  and \textit{Sentence Transformer Model}~\cite{sentence_transformer} as embeddings. For the embedding storage  an in-memory vector store is used. The vector store is populated with chunks of the external-supplied documents. Chunks have a size of 1000 characters each with an overlap of 200 characters.

The choice of the LLM was made by comparing three different models as detailed in Section~\ref{sec:results}. The MMR metric was selected because it yielded better results with respect to the cosine similarity metric~\cite{cosine_similarity} initially used which frequently returned the same chunk multiple times, thereby diminishing the specific knowledge provided to the LLM.

\toolname{} has been implemented by using LangChain~\cite{LangChain}, a framework for the development of applications powered by LLMs, and its extension LangGraph~\cite{LangGraph}. LangChain implements a standard interface not only for LLMs but also for related technologies, such as embedding models and vector stores of hundreds of providers. LangGraph is designed to build robust and stateful multi-actor applications with LLMs by modeling steps as edges and nodes in a graph. LangChain has been chosen because of its popularity in the scientific community and because of its powerful agent-creation libraries. Beyond agents with well-defined high-level interfaces, LangGraph also supports creating agents backed by a low-level, highly controllable API, enabling deep customization of agent logic.

By using the pre-built {\it ReAct agent constructor} we implemented the retrieval and generation steps of the \textit{RAG} subsystem as a call to external tools. Document chunks, retrieved from the vector store, are incorporated into the message sequence sent to the LLM just as if they were messages obtained from  invoking any other external tool. This approach offers the advantage that the model itself generates queries for the retrieval phase, rewriting user messages into more effective search queries. Furthermore, it fully automates direct responses that do not require a retrieval phase, such as responses to generic user greetings.

\section{A real case study: the CyRIS framework}
\label{sec:casestudy}
CyRIS (Cyber Range Instantiation System) \cite{CyRIS} is an open-source tool that provides a flexible, scalable, and low-cost mechanism for managing security training environment. CyRIS supports KVM and AWS virtualization for several guest machines OS, including Ubuntu, CentOS and Windows. The CyRIS input file is structured as a YAML document, which is divided into three sections: \texttt{host\_settings}, \texttt{guest\_settings}, and \texttt{clone\_settings} (see Listing~\ref{lst:cyris_file}). It should be noted that CyRIS facilitates the deployment of CR instances to one or more host servers. 

\begin{lstlisting} [caption=Basic CyRIS cyber range description file, label={lst:cyris_file},basicstyle=\ttfamily\scriptsize]
- host_settings:
    id: host_1
    mgmt_addr: localhost
    virbr_addr: 192.168.10.1
    account: user
- guest_settings:
    id: desktop
    basevm_host: host_1
    basevm_config_file: /home/user/images/basevm.xml
    basevm_type: kvm
- clone_settings:
    range_id: 1
    hosts:
      - host_id: host_1
        instance_number: 1
        guests:
          - guest_id: desktop
            number: 1
            entry_point: yes
        topology:
          - type: custom
            networks:
              - name: office
                members: desktop.eth0
\end{lstlisting} 

We chose CyRIS as CR instantiation framework because it is one of the few platforms that can provide security features even at the CR description stage, such as: 
\begin{inparaenum} [(i)]
    \item traffic capture,
    \item the ability to perform attacks and malware emulation,
    \item the possibility to configure firewall.
\end{inparaenum}
In addition, the GitHub repository \cite{cyris_github} is updated and maintained, making it easy to install and use. The main advantage of CyRIS is its detailed documentation not only on the architecture of the framework, but also on the syntax and semantics to be used when writing the input file. This aspect, together with the examples of description files already provided by the authors, is an excellent knowledge base for \toolname{}. 

Specifically, the initial documents provided to \toolname{} as external knowledge included the most recent version of the paper outlining CyRIS, its user guide, and six CR description files. Despite the adequacy of the documentation, an initial phase of manual document analysis was necessary to remove pages containing information irrelevant to the preparation of the description file. The pages removed included those describing the installation of CyRIS and those documenting performance analysis. To make \toolname{} working with CyRIS we used a total of 28 pages of documentation.

Notwithstanding the present limitations of CyRIS with regard to the supported security features, including the emulation of only a few types of attacks and malware, CyRIS emerged as an excellent framework for testing our approach.

\section{Results and discussion}
\label{sec:results}

\newcommand{\totaltests}{20}
\newcommand{\allfeaturetest}{3}
\newcommand{\succesfullytest}{18}
\newcommand{\onesteprequiredtest}{4}
\newcommand{\twosteprequiredtest}{9}
\newcommand{\threesteprequiredtest}{5}
\newcommand{\allsemanticallyincorrecttest}{5}
\newcommand{\mediumseveritysemanticerrortest}{2}
\newcommand{\averagecheckertoolcalling}{2.05}

In this section we evaluate the coverage that \toolname{} offers at the current stage through different methodological strategies. 

First, a comparative analysis was conducted among the outputs generated: (i) by using only an LLM, (ii) through a RAG system, and (iii) by using  \toolname{}. The comparative assessment yielded significant discrepancies in accuracy, completeness, and contextual relevance across the three approaches.

Subsequently, we conducted both qualitative and quantitative analyses  by generating \totaltests{} CR descriptions, of progressively increasing complexity, encompassing all CyRIS characteristics. This  test enabled the measurement of the success rate of \toolname{} across varying difficulty levels and configuration requirements, thereby providing robust evidence of the system's capabilities under diverse operational scenarios.

Additionally, we evaluated \toolname{}'s performance using various tool-calling LLMs from different providers: \textit{Claude 3.7 Sonnet} from Anthropic, \textit{Gpt-4o-mini} from OpenAI~\cite{GPT_4o_mini} and~\textit{Mistral Large} from Mistral AI~\cite{Mistral_large}. All models successfully completed the task, despite the increasing complexity of the CR to be instantiated. However, minor discrepancies among the models were identified, which ultimately guided the selection of \textit{Claude 3.7 Sonnet} as the primary LLM for \toolname{}. 

\subsection{LLM-Specific constraints in CR generation}
In order to adequately interface an LLM asking for CR generation, some specific concerns have to be properly addressed. In particular, the amount of required details in the user prompt and the way external tools are invoked both are both specific to the employed LLM. The former concern can be easily resolved by specifying the additional information needed by the specific LLM in the system prompt, so that the user input can be kept as generic as possible. For example, in the case of Gpt-4o-mini, it was necessary to specify in the system prompt a message suggesting that the model first retrieve as much information as possible about the CR framework chosen by the user. In the case of mistral, it was necessary to specify that each section of the descriptor files for CyRIS begin with the character \verb+'-'+. These specifications are instrumental in reducing the time required to generate the correct file, but more importantly, they prevent the LLM from performing superfluous computations, thus avoiding reaching the token limit. 

The second concern regards the ability of the LLM to independently repeat multiple verification steps on the generated output taking into account the feedback of the  checker tool. In contrast to the Anthropic and OpenAI models, which can perform this task autonomously, the Mistral AI model required the explicit implementation of a corresponding loop in its agent code.

\subsection{\toolname{} vs. LLM and vs. RAG}
We conducted a series of controlled tests across three configurations: the base LLM, the RAG subsystem, and the full system of \toolname{}. Specifically, we wrote the textual descriptions of $10$ low-complexity CR scenarios and submitted them to each of the three tool configurations. Then we collected, analyzed, and compared the outputs produced. During the test involving only the base LLM, additional details regarding the CyRIS framework syntax were explicitly provided to the model to ensure a fairer comparison. The results obtained are summarized in Table~\ref{tab:system-comparison}.
\vskip-20pt
\begin{table}[ht]
\centering
\caption{Performance Comparison of System Configurations for CR Generation}
\label{tab:system-comparison}
{\scriptsize
\begin{tabular}{|l||p{3cm}|p{3cm}|p{3cm}|}
\hline
\textbf{Metric} & \textbf{Base LLM} & \textbf{RAG} & \textbf{\toolname{}} \\
\hline
\hline
Successful tests & 0/10 & 6/10 & 10/10 \\
\hline
Failure reasons & Lack of framework knowledge despite details provided & Incomplete user requirements \newline Syntax errors & \multicolumn{1}{c|}{-} \\
\hline
Key capabilities & \multicolumn{1}{c|}{-} & Specific CR platform knowledge & RAG capabilities \newline Error correction \newline User interaction \newline Generation of configuration files for self-devised CR scenarios \\
\hline
Main limitations & Ambiguous output \newline Incorrect syntax \newline Failure to meet requirements & Absence of output  verification \newline Limited human interaction \newline Knowledge-base quality & Knowledge-base quality \\
\hline
\end{tabular}
}
\end{table}
\vskip-20pt
The analysis of LLMs outputs revealed a conspicuous deficiency in their capability to generate correct description files, despite the provision of detailed user inputs, making the base LLM incapable of performing the designated tasks. The generated output is often ambiguous and does not align with the specifications of the CR framework stipulated by the user. This outcome underscores the importance for equipping the base LLM with specialized knowledge about the CR framework.

The implementation of a system that is exclusively dependent on RAG signifies a substantial enhancement in the automated generation of description files within a designated CR framework. The integration of knowledge extracted from documents allows the model to produce output that is more closely aligned with the syntax required by the framework, reducing the indeterminacy typical of unconditional generative models.

More in detail, in six out of ten tests, the pure RAG system correctly generated configuration files that conformed to both the framework syntax and the user-specified requirements. In the remaining four tests, the failure was attributed to two main factors: (1)  incomplete or inaccurate user requirements, in one test, and (2) syntax errors,  in the other three tests. In the first case, failures resulted from the user’s omission of mandatory information essential for CR instantiation, such as the \texttt{entry\_point} or the network \texttt{topology} attribute required by the CyRIS CR specification format. In the second case, the errors were caused by formatting issues, such as incorrect indentations in the generated output.

These issues underscore the inherent limitation of using a pure RAG system. These limitations can be effectively mitigated through the use of Agentic RAG techniques, which introduce mechanisms to automatically verify and correct the LLM generated output at each step.

The Agentic RAG-based approach of \toolname{} plays a crucial role in reducing the number of cases in which the system fails to meet user's requests. The LLM can interact directly with a remote server, thereby automatically launching the CR framework and requesting the CR instantiation based on the generated input file. Moreover, using the error messages returned by the framework, \toolname{} successfully passed tests that had previously failed due to syntax errors.

Furthermore, by leveraging the agent's capacity for iterative interaction with the user, a human-in-the-loop approach can be employed to address the test failed in case (1). In these situations, the agent could ask the user if the missing mandatory parameters should be automatically assigned according to reasonable criteria or if the user would prefer to specify them manually. Adding this interaction enabled the successful generation of a correct output also in this case.

\subsection{Quantitative and qualitative analysis}
In order to assess the current capabilities and limitations of \toolname{}, \totaltests{} textual descriptions of training scenarios, of increasing complexity, were manually written by domain experts. Such descriptions were used as prompts to \toolname{}  requesting it to instantiate them on CyRIS.
In writing these scenarios, the experts took care to  include of all features supported by CyRIS-based CRs. This  was done to evaluate \toolname{} in the context of different potential user requirements. Furthermore, at least \allfeaturetest{} of these  scenarios were purposely devised to include all the features supported by CyRIS, in order to evaluate the behavior of \toolname{} under the most complex descriptions.

The tests were carried out using the configuration described in Section~\ref{sec:ragagents} using basic subscription accounts to  Claude 3.7 Sonnet. The evaluation process entailed a maximum of three retry attempts, with a test considered to be correctly completed if successfully passed within these bounds. Performance evaluation was conducted considering three main factors:
\begin{enumerate}
    \item \textit{Correct execution of the required task}: the ability to correctly complete the assigned task, which could consist solely of configuration file generation or even automatic deployment to a remote server.
    \item \textit{Number of iterations required for completion}: the number of iterations required to correct any errors in the intermediate output.
    \item \textit{Semantic validity of the generated output}: since the verification performed by CyRIS via tool calling ensures only the syntactic correctness of the generated file, domain experts manually checked the semantic consistency. They examined the configuration files and their instantiated CRs to determine whether the output was in accordance with the user's requests.
\end{enumerate}

\begin{figure}[ht]
    \centering
    \includegraphics[width=0.75\linewidth]{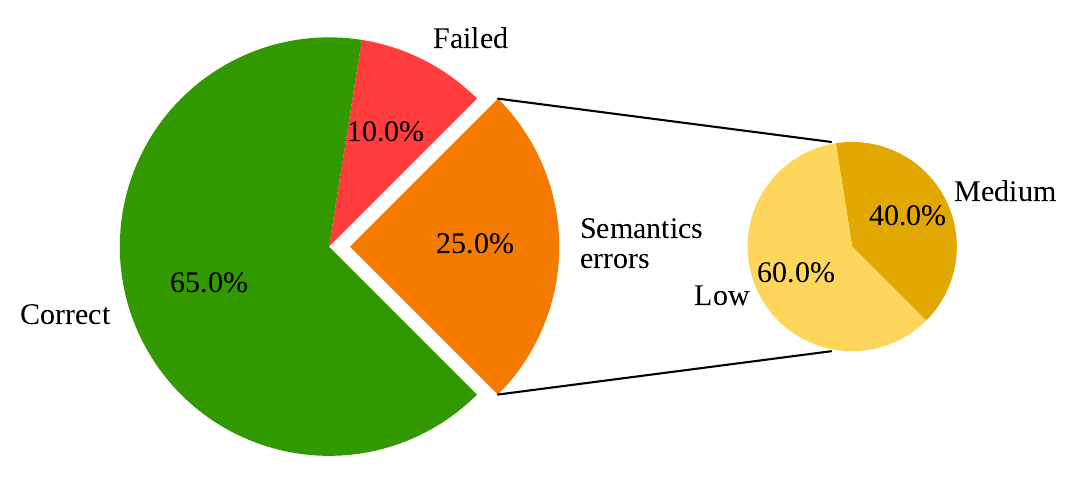}
    \caption{Performance analysis}
    \label{fig:pie_of_pie}
\end{figure}
The results obtained are presented in Fig.~\ref{fig:pie_of_pie}.
An in-depth analysis of the results yielded the following observations. Of the \succesfullytest{} tasks successfully completed by \toolname{}, only \onesteprequiredtest{} instances were solved in one \toolname{} iteration (see Section~\ref{sec:schema} for more details about \toolname{} performed iterations). It should be noted that these \onesteprequiredtest{} instances were the same that would have been correctly addressed also  by the pure RAG configuration. For the other instances the initial output contained syntactic errors that necessitated agent iterations to be corrected, see Table~\ref{tab:test_summary}.

These results corroborate the observations made in the preliminary study and further underscore the need for an Agentic RAG-based approach. Indeed, instances that were successfully completed on the initial attempt correspond to low-complexity scenarios, such as those used in the preliminary tests. However, as the complexity of the user request increased, a pure RAG system proved incapable of completing the task correctly. Of the \allfeaturetest{} tests that included all CyRIS features, 2 were successfully completed. The failures observed in the remaining case was mainly due to the \textit{token limit constraints} associated with LLM usage, which prevented the complete generation of the required configuration files.
\begin{table}[ht]
    \centering
    \caption{Summary of \toolname{} test results}
    \label{tab:test_summary}
    \begin{tabular}{|lcc|}
        \hline
        \textbf{Test category} & \textbf{Count} & \textbf{Percentage (\%)} \\
        \hline
        \hline
        
        Syntactic successfully tests & {\bf \succesfullytest} & {\bf  \the\numexpr100*\succesfullytest/\totaltests\relax\%} \\
        
        \quad Completed in one iteration & \quad \onesteprequiredtest & \quad\ 20\% \\
        \quad Completed in two iterations & \quad\twosteprequiredtest\relax & \quad 45\% \\
        \quad Completed in three iterations & \quad\threesteprequiredtest\relax & \quad 25\% \\
        \hline
        
        Failed tests & {\bf \the\numexpr\totaltests-\succesfullytest\relax} & {\bf \the\numexpr100*(\totaltests-\succesfullytest)/\totaltests\relax\% }\\
        \hline
        Total & \totaltests & 100\% \\
        \hline
    \end{tabular}
\end{table}
A secondary consideration pertains to the severity of the identified semantic errors. We identified three categories of errors as follows:
\begin{enumerate} [(i)]
    \item \textit{High-Severity errors}: the generated configuration does not include all the virtual machines requested by the user and/or all the physical hosts required for deployment;
    \item \textit{Medium-Severity Errors}: failure in the emulation set up of important characteristics of the scenario such as attacks or malware emulation;
    \item \textit{Low-Severity errors}: failures that result in the incorrect set up of non-relevant characteristics of the scenario such as alteration of the username on a VM or the failure to execute a program that is of marginal relevance to the scenario under investigation.
\end{enumerate}

The analysis revealed that of the \allsemanticallyincorrecttest{} tests in which domain experts detected semantic errors, none were of the highest severity, \mediumseveritysemanticerrortest{} errors were classified as medium-severe, while the others 3 had minimal impact. This substantiates the reliability of the employed approach (see Table~\ref{tab:test_summary2}). 
\vskip-20pt
\begin{table}[ht]
    \centering
    \caption{Completed tests with semantic errors}
    \label{tab:test_summary2}
       \begin{tabular}{|lc|} 
        \hline
        High severity errors & \quad \textbf{0}\\
        \hline
        Medium severity errors & \quad \textbf{\mediumseveritysemanticerrortest}\\
        \quad Completed in two iterations & \quad\quad 1\\
        \quad Completed in three iterations & \quad\quad 1\\
        \hline
        Low severity errors & \textbf{\quad\the\numexpr\allsemanticallyincorrecttest-\mediumseveritysemanticerrortest\relax} \\
        \quad Completed in two iterations & \quad\quad 2\\
        \quad Completed in three iterations & \quad\quad 1\\
        \hline
        Total & \quad \textbf{\allsemanticallyincorrecttest}\\
        \hline
    \end{tabular}
\end{table}
\vskip-20pt
In all the tests the semantic correctness of the output was evaluated just after the agent has completed the task within a maximum of three attempts. Nevertheless, it is important to highlight that any identified semantic errors could have been easily rectified by a user with basic familiarity with the CR platform. This can be accomplished through interaction with the agent, where the user can issue an additional request to correct the generated configuration. Such corrections are facilitated by the memory mechanism embedded within \toolname{} architecture, which enables the system to maintain context and adapt to the user’s requests. 

Finally, a notable consideration is related to the distinctive feature of \toolname{} to autonomously conceive a simulation scenario and generate the corresponding configuration file without explicit user specification. Given a prompt requesting the creation of a CR based on a specific platform (e.g., CyRIS), \toolname{} was able to ideate a meaningful scenario, reasonably configure the virtual machines (e.g., installing appropriate programs and executing coherent scripts), and sensibly connect them within a network. This level of autonomy marks a significant advancement over previous approaches based on random scenario generation, offering platform-adapted outputs.

\section{Related work}
\label{sec:related}
Among the proposed CR systems, several have gained prominence. CRATE (Cyber Range and Training Environment)~\cite{Crate_paper} is a VM-based emulation type CR operated by the Swedish Defence Research Agency (FOI). CRATE instantiates the machines from a JSON description of the configurations to be emulated. Additionally, graphical support is provided to simplify scenario definition and access to the CR for users and instructors. 

Subsequent improvements in CR portability and scenario sharing were proposed with Nautilus~\cite{Nautilus_framework}, a CR platform that provides a training environment along with a {\it marketplace} platform allowing to share scenarios, scripts or other pre-implemented vulnerabilities and CVEs. Nautilus leverages cloud technologies to semi-automate deployment of vulnerable systems. Furthermore, it provides a graphical interface to initialize/terminate a training scenario through a remote virtual console available in the Nautilus Web Interface. Real-life scenarios can be written either in a custom Scenario Definition Language or by using the web interface.

A novel lightweight framework for CR orchestration is CyExec*~\cite{CyExec*_CR}, a Docker based CR that encompasses a system that automatically generates multiple scenarios with the same learning objectives utilizing DAG (Directed Acyclic Graph)-based scenario randomization. It leverages container-type virtualization, which offers a lightweight execution environment to run multiple virtual instances efficiently and reducing overall costs using the power of  dockerfiles and {\tt docker-compose} for topology generation.

Further advancements in CR verification were introduced in~\cite{VSDL_CR}, where the authors are the first to propose a method to formally verify the noncontradictory of the scenario. Their framework relies on the virtual scenario description language (VSDL), a domain-specific language for defining high-level features of the desired infrastructure while hiding low-level details. The VSDL specification is then converted into an SMT problem. If this problem is found to be satisfiable, a model is returned that can be used to create the infrastructure.

Despite these advances, many of the proposed solutions have proven viable only for specific CR frameworks for which they were designed. Therefore, \toolname{} is the first Agentic RAG for the configuration and deployment of CRs compatible with multiple CR framework. This implies that \toolname{} can be employed to generate CRs based on different frameworks, thus enabling instructors to leverage the specific advantages offered by each platform.

The present study is situated within the emerging body of research that explores the application of LLMs agents in the domain of cybersecurity. Specifically, our approach aligns with recent studies that leverage the reasoning and generation capabilities of LLM agents for threat intelligence, vulnerability detection, malware and anomaly detection, fuzz and program repair, LLM assisted attack and (in)secure code generation, as reported in~\cite{LLM_for_CS_syrvey}.

In~\cite{LLM_hack_website} the authors show that LLM agents can autonomously hack websites, performing tasks as complex as blind database schema extraction and SQL injections without human feedback and without the need to know the vulnerability beforehand. This capability is uniquely enabled by the use of the tool and leveraging the extended context.
Similar is the work presented in~\cite{LLM_1day_vuln} where the authors presented that LLM agents powered by GPT-4 can autonomously exploit one-day vulnerabilities in real-world systems given the CVE description.

The use of multi-agent systems for cybersecurity goals is explored using SecurityBot~\cite{LLM_agents_with_RL_agents}, a mechanism to enable effective collaboration between LLM agents and pre-trained RL agents that supports cybersecurity operations for red-team and blue-team tasks.

\section{Conclusion and future works}
\label{sec:concl}
In this work \toolname{} has been presented as a Agentic RAG-based solution for the automatic generation of CR configuration files, and the subsequent deployment, starting from a textual description, in a natural language, of the desired scenario. \toolname{} is designed to be independent of any specific CR platform and can interact with different CR frameworks by simply adapting the set of documents provided as augmented knowledge to the RAG component. \toolname{} ensures significantly better performance compared to a pure RAG approach, while maintaining the same level of system complexity. Furthermore, \toolname{} successfully completed the generation of 90\% of the tested simulation scenarios, confirming the validity and effectiveness of the proposed approach. 

The most relevant current limitation is that \toolname{} cannot determine a priori whether the user's request is within the capabilities of the CR framework. To illustrate this point, consider the case of CyRIS. The system is unable to independently detect whether the user is requesting an unsupported network topology, such as a configuration other than the only one currently implemented bus topology. This failure can result in attempts to generate configuration files that can lead to subsequent failures during the CR instantiation phase. 
To address this challenge, a potential future development involves integrating a mechanism for prior validation of the user's request. This mechanism would be based on structured knowledge of the target framework, thereby enabling the agent to identify any inconsistencies prior to output generation. Following this identification, the agent would then communicate the inconsistencies to the user. 

A further extension for future development involves enhancing the Agentic RAG to enable it to correct semantic errors within the generated configuration. This would provide a more comprehensive solution by allowing the system to autonomously address issues related to the configuration that do not adhere to the user's specifications.


Another important direction is to test our approach using open-source LLMs, with the goal of identifying the smallest model within a family of models that supports RAG interaction and tool invocation, yet is capable of correctly completing the tasks. By ``smallest model'', we refer to the model with the least dimensionality. This development aims to demonstrate that the Agentic RAG approach incurs minimal cost while maintaining optimal performance, further validating the efficiency and scalability of the system.

\section*{Acknowledgment}
This work was partially supported by the SERICS project (PE00000014) under the MUR National Recovery and Resilience Plan funded by the European Union - NextGenerationEU.\\
The work of Francesco A. Pironti was supported by Agenzia per la cybersicurezza nazionale under the 2024-2025 funding program for promotion of XL cycle PhD research in cybersecurity (CUP H23C24000640005).



%

\bibliographystyle{IEEEtran}

\bibliography{IEEEabrv,refs}

\begin{thebibliography}{10}
\providecommand{\url}[1]{#1}
\csname url@samestyle\endcsname
\providecommand{\newblock}{\relax}
\providecommand{\bibinfo}[2]{#2}
\providecommand{\BIBentrySTDinterwordspacing}{\spaceskip=0pt\relax}
\providecommand{\BIBentryALTinterwordstretchfactor}{4}
\providecommand{\BIBentryALTinterwordspacing}{\spaceskip=\fontdimen2\font plus
\BIBentryALTinterwordstretchfactor\fontdimen3\font minus
  \fontdimen4\font\relax}
\providecommand{\BIBforeignlanguage}[2]{{%
\expandafter\ifx\csname l@#1\endcsname\relax
\typeout{** WARNING: IEEEtran.bst: No hyphenation pattern has been}%
\typeout{** loaded for the language `#1'. Using the pattern for}%
\typeout{** the default language instead.}%
\else
\language=\csname l@#1\endcsname
\fi
#2}}
\providecommand{\BIBdecl}{\relax}
\BIBdecl

\bibitem{Grimaldi_2024}
\BIBentryALTinterwordspacing
A.~Grimaldi, J.~Ribiollet, P.~Nespoli, and J.~Garcia-Alfaro,
  ``\BIBforeignlanguage{en}{Toward next-generation cyber range: A comparative
  study of training platforms},'' in \emph{\BIBforeignlanguage{en}{Lecture
  Notes in Computer Science}}.\hskip 1em plus 0.5em minus 0.4em\relax Springer
  Nature Switzerland, 2024, pp. 271--290. [Online]. Available:
  \url{http://dx.doi.org/10.1007/978-3-031-54129-2_16}
\BIBentrySTDinterwordspacing

\bibitem{Crate_paper}
\BIBentryALTinterwordspacing
T.~Gustafsson and J.~Almroth, ``\BIBforeignlanguage{en}{Cyber range automation
  overview with a case study of {CRATE}},'' in
  \emph{\BIBforeignlanguage{en}{Lecture Notes in Computer Science}}.\hskip 1em
  plus 0.5em minus 0.4em\relax Springer International Publishing, 2021, pp.
  192--209. [Online]. Available:
  \url{http://dx.doi.org/10.1007/978-3-030-70852-8_12}
\BIBentrySTDinterwordspacing

\bibitem{OpenStack_based_CR}
\BIBentryALTinterwordspacing
S.~Zhou, J.~He, T.~Li, X.~Lan, Y.~Wang, H.~Zhao, and Y.~Li, ``Automating the
  deployment of cyber range with openstack,'' \emph{The Computer Journal},
  vol.~67, pp. 851--863, 2023. [Online]. Available:
  \url{http://dx.doi.org/10.1093/comjnl/bxad024}
\BIBentrySTDinterwordspacing

\bibitem{LLM_for_CR}
M.~Mudassar~Yamin, E.~Hashmi, M.~Ullah, and B.~Katt, ``Applications of llms for
  generating cyber security exercise scenarios,'' \emph{IEEE Access}, vol.~12,
  pp. 143\,806--143\,822, 2024.

\bibitem{Singh2025}
\BIBentryALTinterwordspacing
A.~Singh, A.~Ehtesham, S.~Kumar, and T.~T. Khoei, ``Agentic retrieval-augmented
  generation: A survey on agentic rag,'' 2025. [Online]. Available:
  \url{https://dx.doi.org/10.48550/ARXIV.2501.09136}
\BIBentrySTDinterwordspacing

\bibitem{CyExec*_CR}
R.~Nakata and A.~Otsuka, ``Cyexec*: A high-performance container-based cyber
  range with scenario randomization,'' \emph{IEEE Access}, vol.~9, pp.
  109\,095--109\,114, 2021.

\bibitem{CyRIS}
R.~Beuran, C.~Pham, D.~Tang, K.-i. Chinen, Y.~Tan, and Y.~Shinoda,
  ``Cybersecurity education and training support system: Cyris,'' \emph{IEICE
  Transactions on Information and Systems}, vol. E101.D, no.~3, pp. 740--749,
  2018.

\bibitem{NIST_CR_1}
\BIBentryALTinterwordspacing
NIST, ``Cyber ranges,'' 2018. [Online]. Available:
  \url{https://www.nist.gov/system/files/documents/2018/02/13/cyber_ranges.pdf}
\BIBentrySTDinterwordspacing

\bibitem{blefari2024towardlogbased}
\BIBentryALTinterwordspacing
F.~Blefari, F.~A. Pironti, and A.~Furfaro, ``Toward a log-based anomaly
  detection system for cyber range platforms,'' in \emph{Proceedings of the
  19th International Conference on Availability, Reliability and Security},
  ser. ARES '24.\hskip 1em plus 0.5em minus 0.4em\relax New York, NY, USA:
  Association for Computing Machinery, 2024. [Online]. Available:
  \url{https://doi.org/10.1145/3664476.3669976}
\BIBentrySTDinterwordspacing

\bibitem{romeo2025unvealing}
F.~Romeo, F.~Blefari, F.~A. Pironti, and A.~Furfaro, ``Unveiling attack
  patterns from {CTF} network logs with process mining techniques,'' in
  \emph{Proceedings of the Joint National Conference on Cybersecurity (ITASEC
  \& SERICS 2025)}, 2025.

\bibitem{cybsecguide:2023}
H.~Taylor, ``What is a cyber range? learn hands-on cybersecurity skills,''
  \url{https://cybersecurityguide.org/resources/cyber-ranges/}, 2023.

\bibitem{NIST_CR_2}
\BIBentryALTinterwordspacing
NIST, ``The cyber range: A guide,'' 2023. [Online]. Available:
  \url{https://www.nist.gov/system/files/documents/2023/09/29/The%20Cyber%20Range_A%20Guide.pdf}
\BIBentrySTDinterwordspacing

\bibitem{NCR_USA}
B.~Ferguson, A.~Tall, and D.~Olsen, ``National cyber range overview,'' in
  \emph{2014 IEEE Military Communications Conference}, 2014, pp. 123--128.

\bibitem{FIRE_paper}
\BIBentryALTinterwordspacing
A.~Gavras, A.~Karila, S.~Fdida, M.~May, and M.~Potts, ``Future internet
  research and experimentation,'' \emph{ACM SIGCOMM Computer Communication
  Review}, vol.~37, pp. 89--92, 2007. [Online]. Available:
  \url{http://dx.doi.org/10.1145/1273445.1273460}
\BIBentrySTDinterwordspacing

\bibitem{wooldridge:2009}
M.~Wooldridge, \emph{An Introduction to MultiAgent Systems}, 2nd~ed.\hskip 1em
  plus 0.5em minus 0.4em\relax Wiley, 2009.

\bibitem{LLM_Agents_survey}
Z.~Xi, W.~Chen, X.~Guo, W.~He, Y.~Ding, B.~Hong, M.~Zhang, J.~Wang, S.~Jin,
  E.~Zhou, R.~Zheng, X.~Fan, X.~Wang, L.~Xiong, Y.~Zhou, W.~Wang, C.~Jiang,
  Y.~Zou, X.~Liu, Z.~Yin, S.~Dou, R.~Weng, W.~Qin, Y.~Zheng, X.~Qiu, X.~Huang,
  Q.~Zhang, and T.~Gui, ``The rise and potential of large language model based
  agents: a survey,'' \emph{Science China Information Sciences}, vol.~68, 2025.

\bibitem{RAG}
P.~Lewis, E.~Perez, A.~Piktus, F.~Petroni, V.~Karpukhin, N.~Goyal,
  H.~K{\"u}ttler, M.~Lewis, W.-t. Yih, T.~Rockt{\"a}schel \emph{et~al.},
  ``Retrieval-augmented generation for knowledge-intensive nlp tasks,''
  \emph{Advances in neural information processing systems}, vol.~33, pp.
  9459--9474, 2020.

\bibitem{LangChain}
\BIBentryALTinterwordspacing
H.~Chase, ``Langchain,'' October 2022. [Online]. Available:
  \url{https://github.com/langchain-ai/langchain}
\BIBentrySTDinterwordspacing

\bibitem{LlamaIndex}
\BIBentryALTinterwordspacing
J.~Liu, ``Llamaindex,'' November 2022. [Online]. Available:
  \url{https://github.com/jerryjliu/llama\_index}
\BIBentrySTDinterwordspacing

\bibitem{Langdroid}
\BIBentryALTinterwordspacing
P.~Chalasani and S.~Jha, ``Langdroid.'' [Online]. Available:
  \url{https://github.com/langroid/langroid}
\BIBentrySTDinterwordspacing

\bibitem{YAML_site}
\BIBentryALTinterwordspacing
C.~E. Ingy~döt Net and O.~Ben-Kiki., ``Yaml,'' 2001. [Online]. Available:
  \url{https://yaml.org/about.html}
\BIBentrySTDinterwordspacing

\bibitem{MMR_metric}
\BIBentryALTinterwordspacing
J.~Goldstein and J.~Carbonell, ``Summarization: (1) using {MMR} for diversity-
  based reranking and (2) evaluating summaries,'' in \emph{TIPSTER TEXT PROGRAM
  PHASE III: Proceedings of a Workshop held at Baltimore, {M}aryland, October
  13-15, 1998}.\hskip 1em plus 0.5em minus 0.4em\relax Association for
  Computational Linguistics, 1998, pp. 181--195. [Online]. Available:
  \url{https://aclanthology.org/X98-1025/}
\BIBentrySTDinterwordspacing

\bibitem{Claude_3.7}
\BIBentryALTinterwordspacing
Anthropic, ``Claude 3.7 sonnet system card,'' 2025. [Online]. Available:
  \url{https://assets.anthropic.com/m/785e231869ea8b3b/original/claude-3-7-sonnet-system-card.pdf}
\BIBentrySTDinterwordspacing

\bibitem{sentence_transformer}
N.~Reimers and I.~Gurevych, ``Sentence-{BERT}: Sentence embeddings using
  {S}iamese {BERT}-networks,'' in \emph{Proceedings of the 2019 Conference on
  Empirical Methods in Natural Language Processing}.\hskip 1em plus 0.5em minus
  0.4em\relax Association for Computational Linguistics, 11 2019, pp.
  3982--3992.

\bibitem{cosine_similarity}
H.~Steck, C.~Ekanadham, and N.~Kallus, ``Is cosine-similarity of embeddings
  really about similarity?'' in \emph{Proc. of WWW '24: The ACM Web Conference
  2024}.\hskip 1em plus 0.5em minus 0.4em\relax ACM, 2024, pp. 887--890.

\bibitem{LangGraph}
\BIBentryALTinterwordspacing
C.~Nuno, B.~Vadym, and F.~William, ``{LangGraph}.'' [Online]. Available:
  \url{https://github.com/langchain-ai/langgraph}
\BIBentrySTDinterwordspacing

\bibitem{cyris_github}
\BIBentryALTinterwordspacing
R.~Beuran, ``cyb3rlab/cyris.'' [Online]. Available:
  \url{https://github.com/cyb3rlab/cyris}
\BIBentrySTDinterwordspacing

\bibitem{GPT_4o_mini}
\BIBentryALTinterwordspacing
O.~2024, ``Gpt-4o system card,'' 2024. [Online]. Available:
  \url{https://arxiv.org/abs/2410.21276}
\BIBentrySTDinterwordspacing

\bibitem{Mistral_large}
\BIBentryALTinterwordspacing
M.~A. team, ``Mistral large 2,'' 2024. [Online]. Available:
  \url{https://mistral.ai/news/mistral-large-2407}
\BIBentrySTDinterwordspacing

\bibitem{Nautilus_framework}
G.~Bernardinetti, S.~Iafrate, and G.~Bianchi, ``Nautilus: A tool for automated
  deployment and sharing of cyber range scenarios,'' \emph{Proceedings of the
  16th International Conference on Availability, Reliability and Security}, pp.
  1--7, 2021.

\bibitem{VSDL_CR}
G.~Costa, E.~Russo, and A.~Armando, ``Automating the generation of cyber range
  virtual scenarios with vsdl,'' \emph{arXiv preprint arXiv:2001.06681}, 2020.

\bibitem{LLM_for_CS_syrvey}
\BIBentryALTinterwordspacing
J.~Zhang, H.~Bu, H.~Wen, Y.~Liu, H.~Fei, R.~Xi, L.~Li, Y.~Yang, H.~Zhu, and
  D.~Meng, ``When {LLM}s meet cybersecurity: a systematic literature review,''
  \emph{Cybersecurity}, 2025. [Online]. Available:
  \url{http://dx.doi.org/10.1186/s42400-025-00361-w}
\BIBentrySTDinterwordspacing

\bibitem{LLM_hack_website}
\BIBentryALTinterwordspacing
R.~Fang, R.~Bindu, A.~Gupta, Q.~Zhan, and D.~Kang, ``{LLM} agents can
  autonomously hack websites,'' \emph{{arXiv}}, 2024. [Online]. Available:
  \url{https://arxiv.org/abs/2402.06664}
\BIBentrySTDinterwordspacing

\bibitem{LLM_1day_vuln}
R.~Fang, R.~Bindu, A.~Gupta, and D.~Kang, ``{LLM} agents can autonomously
  exploit one-day vulnerabilities,'' \emph{arXiv preprint arXiv:2404.08144},
  2024.

\bibitem{LLM_agents_with_RL_agents}
Y.~Yan, Y.~Zhang, and K.~Huang, ``Depending on yourself when you should:
  Mentoring {LLM} with {RL} agents to become the master in cybersecurity
  games,'' \emph{arXiv preprint arXiv:2403.17674}, 2024.

\end{thebibliography}

\end{document}